\documentclass[prb,preprint,showpacs,preprintnumbers,amsmath,amssymb]{revtex4}

\usepackage{graphicx}
\usepackage{dcolumn}
\usepackage{bm}

\begin{document}

\newcommand*{\cm}{cm$^{-1}$\,}
\newcommand*{\Tc}{T$_c$\,}


\title{Transport and anisotropy in single-crystalline SrFe$_2$As$_2$ and
 $A_{0.6}$K$_{0.4}$Fe$_2$As$_2$ ($A$ = Sr, Ba) superconductors}
\author{G. F. Chen}
\author{Z. Li}
\author{J. Dong}
\author{G. Li}
\author{W. Z. Hu}
\author{X. D. Zhang}
\author{X. H. Song}
\author{P. Zheng}
\author{N. L. Wang}
\author{J. L. Luo}

\affiliation{Beijing National Laboratory for Condensed Matter
Physics, Institute of Physics, Chinese Academy of Sciences,
Beijing 100190, China}


\begin{abstract}
We have successfully grown high quality single crystals of
SrFe$_2$As$_2$ and A$_{0.6}$K$_{0.4}$Fe$_2$As$_2$(A=Sr, Ba) using
flux method. The resistivity, specific heat and Hall coefficient
have been measured. For parent compound SrFe$_2$As$_2$, an
anisotropic resistivity with $\rho_c$/$\rho_{ab}$ as large as 130
is obtained at low temperatures. A sharp drop in both in-plane and
out-plane resistivity due to the SDW instability is observed below
200 K. The angular dependence of in-plane magnetoresistance shows
2-fold symmetry with field rotating within ab plane below SDW
transition temperature. This is consistent with a stripe-type spin
ordering in SDW state. In K doped
A$_{0.6}$K$_{0.4}$Fe$_2$As$_2$(A=Sr. Ba), the SDW instability is
suppressed and the superconductivity appears with T$_c$ above 35
K. The rather low anisotropy in upper critical field between
H$\parallel$ab and H$\parallel$c indicates inter-plane coupling
play an important role in hole doped Fe-based superconductors.

\end{abstract}



\maketitle

The recent discovery of superconductivity with transition
temperature T$_c\sim$26 K in LaFeAsO$_{1-x}$F$_x$ has generated
tremendous interest in scientific community\cite{Kamihara08}.
Shortly after this discovery, the T$_c$ was raised to 41-55 K by
replacing La by rare-earth Ce, Sm, Pr, Nd, etc, making those
systems with T$_c$ exceeding 50 K.\cite{Chen1,XHChen,Ren2,CWang}
The undoped quaternary compounds crystallize in a tetragonal
ZrCusiAs-type structure, which consists of alternate stacking of
edge-sharing Fe$_2$As$_2$ tetrahedral layers and La$_2$O$_2$
tetrahedral layers along c-axis. Very recently, superconductivity
with T$_c$ up to 38 K was discovered in AFe$_2$As$_2$(A=Ba, Sr,
Ca) upon K or Na-doping.\cite{Rotter2, Chen, Wu, Sasmal, Wu2}
AFe$_2$As$_2$ compounds crystallize in a tetragonal
ThCr$_2$Si$_2$-type structure with identical Fe$_2$As$_2$
tetrahedral layers as in LaFeAsO, but separated by single
elemental A-layers. These compounds contain no oxygen in A layers.
The simpler structure of AFe$_2$As$_2$ system makes it more
suitable for research of intrinsic physical properties of Fe-based
compounds.

Except for a relatively high transition temperature, the system
displays many interesting properties. The existence of a
spin-density-wave (SDW) instability in parent LaFeAsO\cite{Dong}
was indicated by specific heat, optical measurements and first
principle calculations, and subsequently confirmed by neutron
scattering,\cite{Cruz} NMR,\cite{Ishida} $\mu$sR,\cite{Klauss} and
M\"{o}ssbaure\cite{Kitao} spectroscopic measurements. The
superconductivity only appears when SDW instability was suppressed
by doping carriers or applying pressure. The competition between
superconductivity and SDW instability was identified in other
rare-earth substituted systems \cite{Chen1,Ding,Chen2}. Besides
the SDW instability, a structural distortion from tetragonal to
monoclinic were also observed for both ReFeAsO(Re=rare earth) and
AFe$_2$As$_2$ (A=Ba, Sr, Ca)\cite{Rotter1, Bao, Ni, Ni3, Ni2,
Tegel}. The structural transition temperature were found to occur
at slightly higher than SDW transition temperature in
LaFeAsO\cite{Cruz}, but the two transitions occur simultaneously
in AFe$_2$As$_2$(A=Ba, Sr, Ca).\cite{Rotter1, Bao, Zhao} The band
structure calculation and neutron scattering experiments indicated
a stripe-type antiforremagnetic structure of Fe moments in SDW
state in LaFeAsO.\cite{Dong, Cruz} In such a spin structure, a
4-fold spin rotation symmetry is broken and reduced to 2-fold. It
is interesting to see if such a 2-fold symmetry could be observed
in angular-dependent magneto-resistance measurements.

In this work, we present resistivity, specific heat and Hall
coefficient measurements on single crystals SrFe$_2$As$_2$ and
A$_{0.6}$K$_{0.4}$Fe$_2$As$_2$ (A=Sr, Ba). A large anisotropic
resistivity with $\rho_c$/$\rho_{ab}$ $\sim$ 130 is observed for
SrFe$_2$As$_2$. Below 200K, a sharp drop in both in-plane and
out-plane resistivity due to the SDW instability is observed,
similar to previously results on polycrystal samples\cite{Chen}.
Moreover, the angular-dependence of in-plane and out-plane
magnetoresistance shows 2-fold symmetry with field rotating within
ab plane below SDW transition temperature. The breaking of 4-fold
rotation symmetry to 2-fold provides transport evidence for the
formation of a stripe-type antiforremagnetic structure of Fe
moments in SDW state. Moreover, the anisotropy of upper critical
field for high quality single crystal
Sr$_{0.6}$K$_{0.4}$Fe$_2$As$_2$ is investigated. The inter-layer
coupling is relatively strong in superconducting samples.

The parent single crystals SrFe$_2$As$_2$ were prepared by the
high temperature solution method using Sn as flux, similar to the
procedure described in ref\cite{Ni3}. The superconducting crystals
A$_{0.6}$K$_{0.4}$Fe$_2$As$_2$ (A=Sr, Ba) were prepared by the
high temperature solution method using FeAs as flux. The starting
materials of Sr or Ba, K, and FeAs in a ratio of 0.5:1:4 were put
into an alumina crucible and sealed in welded Ta crucible under
1.6 atmosphere of argon. The Ta crucible were then sealed in an
evacuated quartz ampoule and heated at 1150 $^{\circ}C$ for 5
hours and cooled slowly to 800 $^{\circ}C$ over 50 hours. The
plate-like crystals with sizes up to 10mm$\times$5m$\times$0.5mm
could be obtained after breaking the alumina crucible. Both
scanning electron microscopy/energy dispersive x-ray (SEM/EDX) and
induction-coupled plasma (ICP) analysis revealed that the
elemental composition of the crystal is
A$_{0.6\pm\delta}$K$_{0.4\mp\delta}$Fe$_2$As$_2$ (A=Sr, Ba) with
$\delta \leq 0.02$. Figure 1(a) shows X-ray diffraction pattern of
parent SrFe$_2$As$_2$ with the 00 $\ell$ reflections. The lattice
constant c = 0.1239 nm was calculated from the higher order peaks,
comparable to that of polycrystalline sample\cite{Chen}. Figure
1(b) shows the EDX analysis of Ba$_{0.6}$K$_{0.4}$Fe$_2$As$_2$
crystal.

The in-plane resistivity was measured by the standard 4-probe
method. The out-plane resistivity was measured using a typical
method for layered materials as shown in inset of Fig. 2(b). The
silver paste was used to cover almost all area of upper and lower
side of the measured sample for two current leads, and leaves two
small holes in the center of both sides for the voltage leads.
Assuming that the current density is uniformly distributed
throughout the cross section, the resistivity can then be measured
with: $\rho_c$ = R S /d, where R is the resistance, S is area of
cross section and d is the thickness of the sample. The error
comes mostly from the measurement of geometry factors. The ac
magnetic susceptibility was measured with a modulation field in
the amplitude of 10 Oe and a frequency of 333 Hz. The Hall
coefficient measurement was done using a five-probe technique. The
specific heat measurement was carried out using a thermal
relaxation calorimeter. All these measurements were preformed down
to 1.8K in a Physical Property Measurement System(PPMS) of Quantum
Design company.

\begin{figure}
\includegraphics[width=7cm,clip]{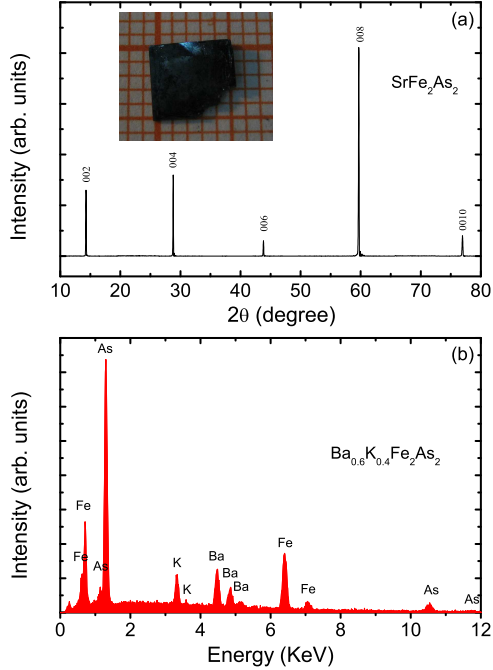}

\caption{(Color online) (a) Single crystal X-ray diffraction
pattern for SrFe$_2$As$_2$. The inset shows the photograph of a
SrFe$_2$As$_2$ single crystal (length scale 1 mm). (b) EDX
analysis of Ba$_{0.6}$K$_{0.4}$Fe$_2$As$_2$ crystal.}
\end{figure}

Figure 2(a) shows the temperature dependence of in-plane
resistivity $\rho_{ab}$ in zero field and 8 T, and Fig. 2(b) shows
the out-plane resistivity $\rho_{c}$ for SrFe$_2$As$_2$ in zero
field. They have similar T-dependent behavior and exhibit a strong
anomaly at about 200 K: the resistivity drops steeply below this
temperature. This is a characteristic feature related to SDW
instability and a structural distortion in parent compounds of
Fe-based high T$_c$ superconductors.\cite{Dong, Cruz} The
transition temperature is slightly lower than that observed in
poly-crystalline SrFe$_2$As$_2$ ($\sim$ 205 K). The anisotropic
resistivity $\rho_c$/$\rho_{ab}$ is found to be $\sim$ 130$\pm$65
at 25 K. It is comparable with that of layered cuprates
YBa$_2$Cu$_3$O$_{7-\delta}$.\cite{Friedmann} From Fig.2(a), we
find that the SDW transition temperature is insensitive to the
applied field, however a large positive magnetoresistance is
observed at low temperatures. At 10 K, the magnetoresistance
[$\rho_{ab}(8T)$-$\rho_{ab}(0T)$]/$\rho_{ab}(0T)$ reaches as high
as 25$\%$. The behavior is similar to that observed in polycrystal
LaFeAsO.\cite{Dong}. The large positive magnetoresistance can be
understood from the suppression of SDW order by applied field and
thus the spin scattering is enhanced.

\begin{figure}
\centerline{\includegraphics[width=7cm]{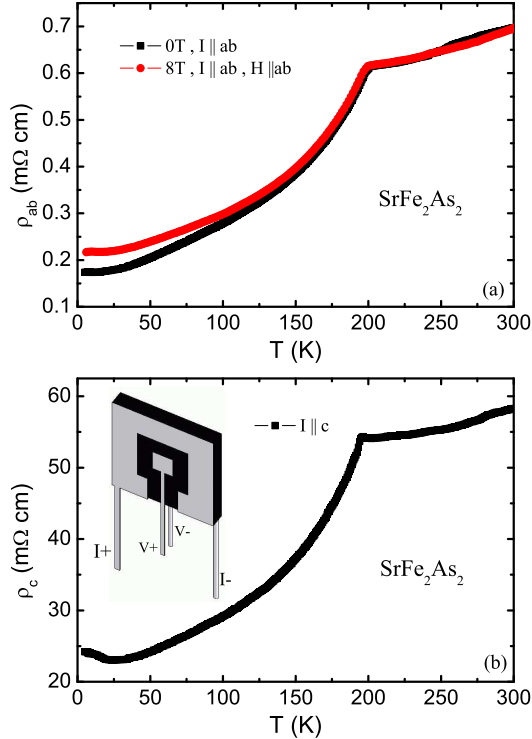}}%
\caption{(Color online) (a) The in-plane resistivity $\rho_{ab}$
for SrFe$_2$As$_2$ in zero field and 8 T, respectively. (b) The
out-plane resistivity $\rho_{c}$ in zero field. The inset shows a
sketch for how to measure $\rho_{c}$.}
\end{figure}

In a system with strong coupling of charge carrier and background
magnetism, angular-dependent-magnetoresistance(AMR) is a useful
tool to detect the magnetic structure of background magnetism. AMR
measurements have been successfully used in understanding the spin
structure of slightly doped high T$_c$ cuprates and charge
ordering Na$_{0.5}$CoO$_2$ systems.\cite{XHChen3,Hu} For this
purpose, we measured the in-plane resistivity with magnetic field
of 10 T rotating within the ab plane for SrFe$_2$As$_2$. Figure 3
shows AMR ($\Delta\rho(\theta)=\rho(\theta)-\rho(90^{\circ})$) at
10, 30, 50, 100, 150 and 200 K, respectively. $\Delta\rho(\theta)$
shows a clear 2-fold oscillation below 200K. However, the 2-fold
oscillation disappears at 200 K, a temperature just above
T$_{SDW}$. Above T$_{SDW}$, the spin orientations of Fe moments
are random so there is no angular dependent magnetoresistance
arising from spin scattering. However below T$_{SDW}$, a large MR
is observed at low temperature and it originates from the enhanced
spin scattering while SDW order is suppressed in external magnetic
field. Therefore the symmetry of the AMR oscillation reflect
directly the symmetry of spin structure. The observation of 2-fold
symmetry oscillations in SrFe$_2$As$_2$ in low temperatures
indicates the stripe-type spin structure of Fe moments in ground
state. The recent neutron diffraction measurements on
SrFe$_2$As$_2$ single crystals \cite{Zhao} and polycrystalline
\cite{Kaneko} indicate that the SDW transition is indeed presented
in SrFe$_2$As$_2$ system below a transition temperature of $\sim$
200-220 K. The stripe-type antiferromagnetic spin structure is
found in the SDW state. Moreover, the angular-dependent
magnetization measurement on BaFe$_2$As$_2$ crystal with field
rotating within the ab plane also show a 2-fold symmetry of
magnetization below SDW transition temperature\cite{Wang},
consistent with the AMR measurements on SrFe$_2$As$_2$ crystals.
Therefore AMR can act as a probe for SDW order in these systems
without having to wait for neutron beam time. Similar 2-fold
symmetry oscillations of AMR has been observed in
Na$_{0.5}$CoO$_2$ which was attributed to ordering of charge
stripes.\cite{Hu}

\begin{figure}
\includegraphics[width=7cm,clip]{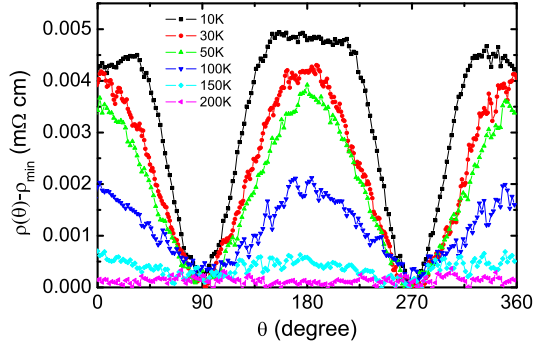}
\caption {(Color online) Angular dependence of in-plane
magnetoresistance($\Delta\rho(\theta)=\rho(\theta)-\rho(90^{\circ})$)
at different temperatures in external magnetic field of 10T
rotating within ab plane for SrFe$_2$As$_2$}
\end{figure}

To get more information about the SDW and structural phase
transition, we preformed specific heat and Hall resistivity
measurement for SrFe$_2$As$_2$. Figure 4(a) shows the temperature
dependence of specific heat C from 2 to 230 K. We can see clearly
a sharp $\delta$-function shape peak at about 200 K with $\Delta$C
$\sim$ 185 J/molK. This is a characteristic feature of a
first-order phase transition. The transition temperature agrees
well with that observed in resistivity measurement. The latent
heat of the transition is estimated being 463$\pm$50 J/mol by
integral the area of the specific heat data around the transition
peak after subtracting the background. The background is estimated
using a linear fit of the specific heat data above and below the
transition. It worth noting that only one peak at around 200 K is
observed in SrFe$_2$As$_2$, different from that of LaFeAsO where
two subsequent peaks at 155 K and 143 K were observed in specific
heat data corresponding to structural and SDW transition,
respectively.\cite{Jin} This suggests the structural transition
and SDW transition occur at same temperature in SrFe$_2$As$_2$,
similar to that observed in BaFe$_2$As$_2$.\cite{Rotter1, Bao} At
low temperatures, a good linear T$^2$ dependence of C/T is
observed indicates that the specific heat C is mainly contributed
by electrons and phonons [see inset Fig. 4(a)]. The fit yields the
electronic coefficient $\gamma$=6.5 mJ/mol.K$^2$ and the Debye
temperature $\theta_D$ = 245 K. Note that the electronic
coefficient is significantly smaller than the values obtained from
the band structure calculations for non-magnetic
state.\cite{Xiang} This can be explained that a partial energy gap
is opened below SDW transition, the smaller experimental value
here could be naturally accounted for by the gap formation which
removes parts of the density of states below the phase transition.
Compared with band calculation, similar smaller electronic
specific heat coefficient was also observed in LaOFeAs due to gap
opening originated from SDW instability.\cite{Dong} In comparison,
we also measured specific heat of superconducting crystal
Sr$_{0.6}$K$_{0.4}$Fe$_2$As$_2$ prepared using FeAs as flux. It is
found that no structural and SDW transition can be observed in
specific heat data. Instead, a specific heat anomaly at a
superconducting transition temperature with T$_c$$\sim$35.6 K is
observed. The specific heat jump $\Delta$C/T$_c$ is found to be
$\sim$ 48 mJ/mol.K$^2$.

\begin{figure}
\includegraphics[width=7cm,clip]{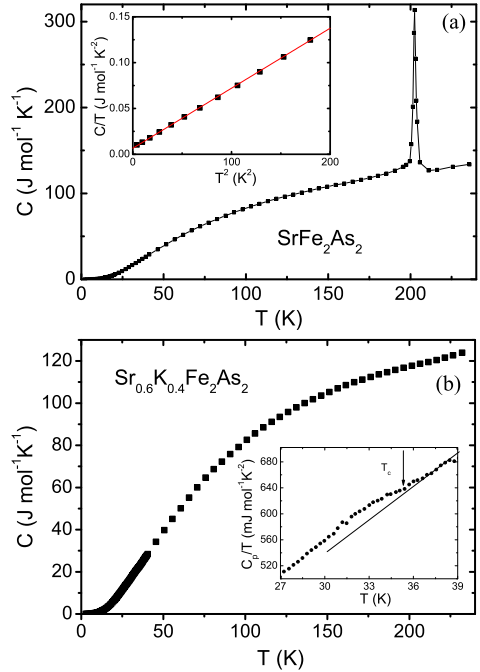}
\caption {(Color online) (a) Temperature dependence of specific
heat C for SrFe$_2$As$_2$. Inset: T$^2$ dependence of C/T in low
temperatures. (b)Temperature dependence of specific heat C for
Sr$_{0.6}$K$_{0.4}$Fe$_2$As$_2$. Inset: C/T around superconducting
transition.}
\end{figure}

Hall coefficient R$_H$ as a function of temperature between 20 and
300 K for SrFe$_2$As$_2$ is shown in Fig. 5. For comparison,
R$_H$(T) for LaFeAsO is also shown in inset of Fig. 5. Above 200
K, the Hall coefficient is negative and nearly temperature
independent for SrFe$_2$As$_2$, indicating conduction carriers are
dominated by electrons. The carrier density is estimated being n =
1.5$\times$10$^{22}$cm$^{-3}$ at 300 K if one-band model is simply
adopted. It is nearly an order higher than that of LaFeAsO with n
$\sim$ 1.8$\times$10$^{21}$cm$^{-3}$ at 300 K obtained by same
method. Optical measurement also indicates a quite large plasma
frequency, $\omega_{p}$ $\sim$ 1.5 eV.\cite{WZHu} The large
carrier number for SrFe$_2$As$_2$ indicate that it is a good
metal. Below 200 K, the Hall coefficient increases slightly to a
positive value, and then drops dramatically to a very large
negative value. The absolute value of R$_H$ at 2 K is about 35
times larger than that at 300 K. The dramatic change reflects the
reconstruction of Fermi surface after SDW transition. The band
calculation show that there are 3 hole pockets around $\Gamma$
point and 2 electron pockets around M point\cite{Dong}. The
experiments seem to indicate that, upon cooling below the SDW
transition temperature, hole pockets are almost fully gapped while
the electron pockets are partially gapped. Therefore, at low
temperatures, the R$_H$ reflect mainly the the un-gapped electron
density around M point, which is significantly small in comparison
with its initial value above 200K.

\begin{figure}
\includegraphics[width=7cm,clip]{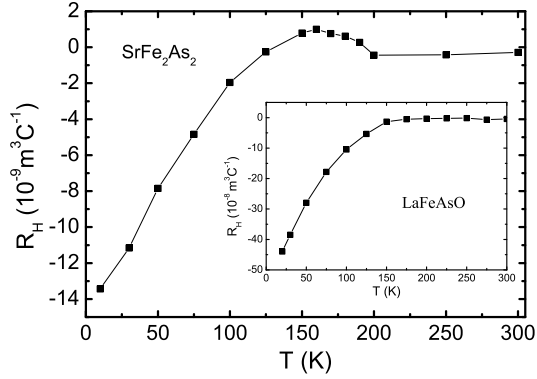}
\caption {(Color online) Temperature dependence of Hall
coefficient for SrFe$_2$As$_2$. For comparison, the temperature
dependence of Hall coefficient of LaFeAsO is shown in the inset.}
\end{figure}

In addition to parent SrFe$_2$As$_2$, the anisotropy of
resistivity and upper critical field for single crystals
A$_{0.6}$K$_0.4$Fe$_2$As$_2$ (A=Sr, Ba) is also investigated.
Figure 6(a) shows the temperature dependence of the in-plane
resistivity in zero field for Sr$_{0.6}$K$_{0.4}$Fe$_2$As$_2$ and
Ba$_{0.6}$K$_{0.4}$Fe$_2$As$_2$ crystals. $\rho_{ab}$ decreases
with decreasing temperature and shows a downward curvature for
Sr$_{0.6}$K$_{0.4}$Fe$_2$As$_2$, consistent with the polycrystal
sample. With further decreasing temperature, an extremely sharp
superconducting transition at 35.5 K with transition width about
0.3 K is observed, indicating high homogeneity of the sample. A
sharp superconducting transition for
Ba$_{0.6}$K$_{0.4}$Fe$_2$As$_2$ is also observed with T$_c$=38 K,
slightly higher than that of Sr$_{0.6}$K$_{0.4}$Fe$_2$As$_2$. The
anisotropic resistivity $\rho_{c}$/$\rho_{ab}$ in normal state for
Sr$_{0.6}$K$_{0.4}$Fe$_2$As$_2$ is found to be 21, much lower than
that of parent SrFe$_2$As$_2$.

\begin{figure}
\includegraphics[width=12cm,clip]{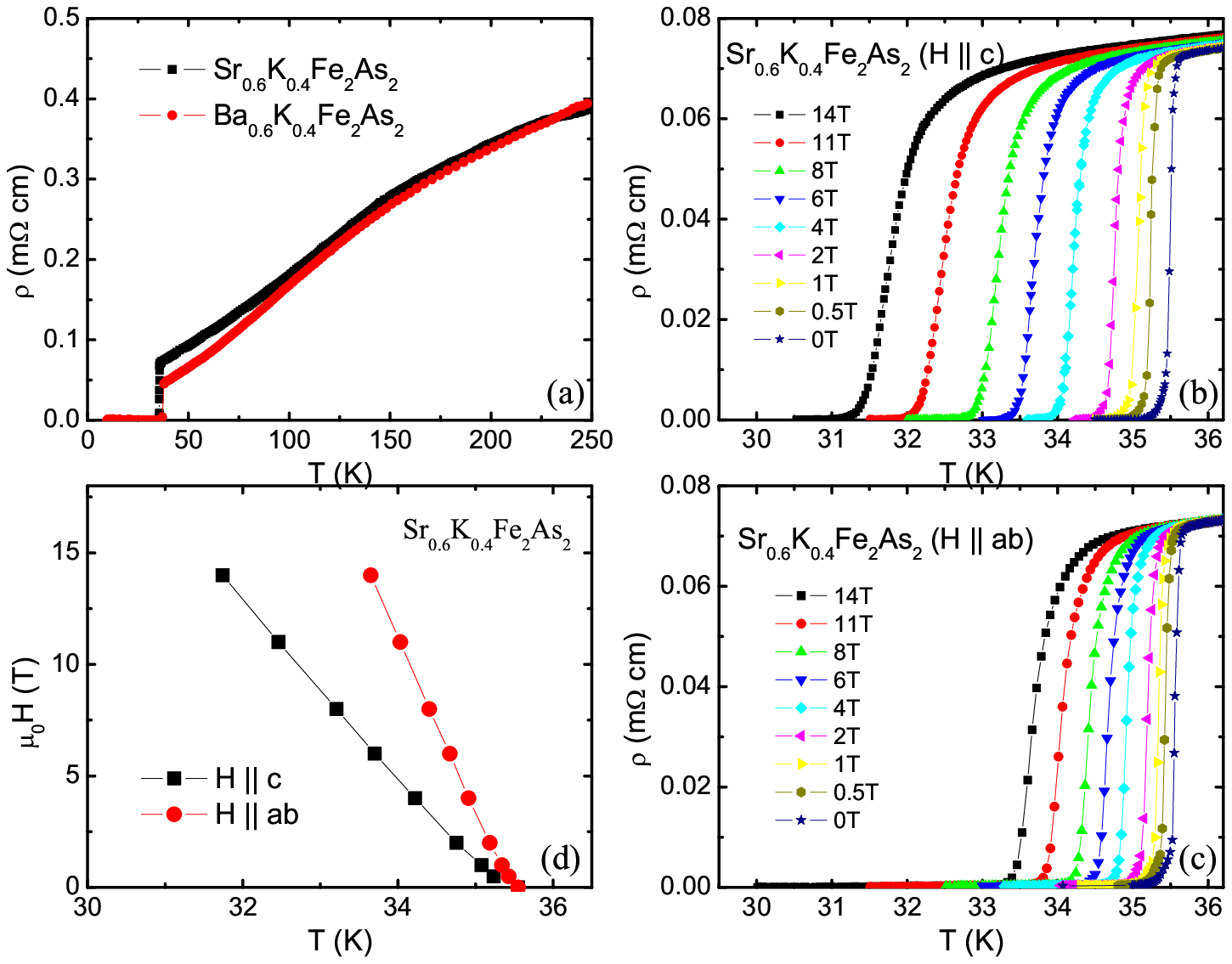}
\caption {(Color online) (a) Temperature dependence of the
in-plane electrical resistivity for
Sr$_{0.6}$K$_{0.4}$Fe$_2$As$_2$ and
Ba$_{0.6}$K$_{0.4}$Fe$_2$As$_2$ in zero field, respectively. (b)
and (c) Temperature dependence of the in-plane electrical
resistivity for Sr$_{0.6}$K$_{0.4}$Fe$_2$As$_2$ at fixed fields up
to 14T for H $\parallel$ c and H $\parallel$ ab, respectively.
(d)H$_{c2}$(T) plot for H $\parallel$ c(closed square) and H
$\parallel$ ab(open circle) , respectively. }
\end{figure}

Figure 6(b) and figure 6(c) show $\rho_{ab}$(T) for
Sr$_{0.6}$K$_{0.4}$Fe$_2$As$_2$ in external magnetic fields up to
14 T along c-axis and within ab plane, respectively. We can see
the superconducting transition is broadened slightly in magnetic
fields up to 14 T. The behavior is different from polycrystalline
LaFeAsO where the superconducting transition is broadened strongly
in magnetic fields.\cite{Chen0} Figure 6(d) shows H$_{c2}$-T$_c$
curves for both H$\parallel$ab and H$\parallel$c, respectively,
where T$_c$ is adopted by a criterion of 90$\%$ of normal state
resistivity. The curves H$_{c2}$(T) are very steep with slopes
-dH$_{c2}^{ab}$ /dT $\mid _{T_c}$=7.57 T/K for H$\parallel$ab and
-dH$_{c2}^{ab}$/dT$\mid_{T_c}$=3.80 T/K for H$\parallel$c. This
indicates the upper critical fields is extremely high for
Sr$_{0.6}$K$_{0.4}$Fe$_2$As$_2$. Using the
Werthamer-Helfand-Hohenberg formula\cite{WHH}
H$_{c2}$(0)=-0.69(dH$_{c2}$/dt)T$_c$ and taking T$_c$=35.5K, the
upper critical fields are: H$_{c2}^{ab}$=185.4 T and
H$_{c2}^{c}$=93.1 T, respectively. The anisotropy ratio
$\gamma$=H$_{c2}^{ab}$/H$_{c2}^{c}$$\approx$2.0.\cite{Ni} The
value of $\gamma$ is close to that of
Ba$_{0.55}$K$_{0.45}$Fe$_2$As$_2$ with $\gamma$ between 2.5 to
3.5. It is lower than that of F-doped NdFeAsO\cite{Wen} with
$\gamma$ $\approx$ 4.3-4.9, and much lower than high T$_c$
cuprates, for example $\gamma$ $\approx$ 7-10 for
YBCO.\cite{Nanda} The lower values of $\rho_{c}$/$\rho_{ab}$ and
$\gamma$ indicate that the inter-plane coupling in
Sr$_{0.6}$K$_{0.4}$Fe$_2$As$_2$ is relative strong.

To summarize, SrFe$_2$As$_2$ and
A$_{0.6}$K$_{0.4}$Fe$_2$As$_2$(A=Sr, Ba) single crystals were
prepared by flux method. The angular dependence of in-plane
resistivity was measured for SrFe$_2$As$_2$. A 2-fold symmetry of
oscillations in AMR is observed at low temperatures which possible
indicate a stripe-type spin structure below SDW temperature. In K
doped A$_{0.6}$K$_{0.4}$Fe$_2$As$_2$(A=Sr. Ba), the SDW
instability is suppressed and instead superconductivity appears
with T$_c$ above 35 K. The upper critical field is rather high
with H$_{c2}^{ab}$=185.4 T, while its anisotropy is rather low.
The inter-plane coupling may play an important role in this
material.

\begin{acknowledgments}
We acknowledge Z. D. Wang, T. Xiang and Z. Fang for helpful
discussions, and Q. M. Meng in experiments. This work is supported
by the National Science Foundation of China, the Knowledge
Innovation Project of the Chinese Academy of Sciences, and the 973
project of the Ministry of Science and Technology of China.

\end{acknowledgments}


\begin{thebibliography}{20}

\bibitem{Kamihara08} Y. Kamihara, T. Watanabe, M. Hirano, and
H. Hosono, J. Am. Chem. Soc. \textbf{130}, 3296 (2008).

\bibitem{Chen1} G. F. Chen, Z. Li, D. Wu, G. Li, W. Z. Hu, J. Dong, P. Zheng, J. L. Luo, and
N. L. Wang,  Phys. Rev. Lett. \textbf{100}, 247002 (2008)

\bibitem{XHChen} X. H. Chen, T. Wu, G. Wu, R. H. Liu, H.
Chen, and D. F. Fang, Nature \textbf{453}, 761 (2008).

\bibitem{Ren2} Z.-A. Ren, J. Yang, W. Lu, W.
Yi, X.-L. Shen, Z.-C. Li, G.-C. Che, X.-L. Dong, L.-L. Sun, F.
Zhou, and Z.-X. Zhao, Europhys. Lett. \textbf{82}, 57002 (2008).

\bibitem{CWang}Cao Wang, Linjun Li, Shun Chi, Zengwei Zhu, Zhi Ren, Yuke Li,
Yuetao Wang, Xiao Lin, Yongkang Luo, Xiangfan Xu, Guanghan Cao,
and Zhu'an Xu, Europhys. Lett. \textbf{83}, 67006 (2008).

\bibitem{Rotter2} M. Rotter, M. Tegel, and D. Johrendt, Phys. Rev.
Lett. \textbf{101}, 107006 (2008).

\bibitem{Chen}G. F. Chen, Z. Li, G. Li, W. Z. Hu, J. Dong, X. D. Zhang, P. Zheng, N. L. Wang, and J. L. Luo, Chinese Phys. Lett. \textbf{25},
3403(2008)

\bibitem{Wu}G. Wu, R. H. Liu, H. Chen, Y. J. Yan, T. Wu, Y. L. Xie, J. J. Ying, X. F. Wang, D. F. Fang, and X. H.
Chen, Europhys. Lett. \textbf{84}, 27010 (2008).

\bibitem{Sasmal}K. Sasmal, B. Lv, B. Lorenz, A. M. Guloy, F. Chen, Y. Y. Xue, and C. W. Chu, Phys. Rev. Lett. \textbf{101}, 107007 (2008)

\bibitem{Wu2}G. Wu, H. Chen, T. Wu, Y. J. Yan, R. H. Liu, X. F. Wang, J. J. Ying, and X. H.
Chen, J. Phys.: Condens. Matt. \textbf{20}, 422201 (2008).

\bibitem{Dong} J. Dong, H. J. Zhang, G. Xu, Z. Li, G. Li, W. Z. Hu, D. Wu,
G. F. Chen, X. Dai, J. L. Luo, Z. Fang, and N. L. Wang, Europhys.
Lett. \textbf{83}, 27006 (2008).

\bibitem{Cruz} C. de la Cruz, Q. Huang, J. W. Lynn,
J. Li, W. R. Ii, J. L. Zarestky, H. A. Mook, G. F. Chen, J. L.
Luo, N. L. Wang, and Pengcheng Dai, Nature \textbf{453},
899(2008).

\bibitem{Ishida} Y. Nakai, K. Ishida, Y. Kamihara, M. Hirano, and Hideo Hosono, J. Phys. Soc. Jpn. \textbf{77},  073701 (2008).

\bibitem{Klauss} H.-H. Klauss, H. Luetkens, R. Klingeler, C. Hess, F. J. Litterst, M. Kraken, M. M. Korshunov,
I. Eremin, S.-L. Drechsler, R. Khasanov, A. Amato, J.
Hamann-Borrero, N. Leps, A. Kondrat, G. Behr, J. Werner, and B.
Buchner, Phys. Rev. Lett. \textbf{101}, 077005 (2008).

\bibitem{Kitao} S. Kitao, Y. Kobayashi, S. Higashitaniguchi, M. Saito, Y. Kamihara, M. Hirano, T. Mitsui, H. Hosono,
and M. Seto, J. Phys. Soc. Jpn. \textbf{77}, 103706 (2008).

\bibitem{Chen2}  G. F. Chen, Z. Li, D. Wu, J. Dong, G. Li, W. Z. Hu, P. Zheng, J. L. Luo,
N. L. Wang, Chin. Phys. Lett. \textbf{25}, 2235 (2008).

\bibitem{Ding} L. Ding, C. He, J. K. Dong, T. Wu, R. H. Liu, X. H. Chen, S. Y. Li, Phys. Rev. B \textbf{77}, 180510(R) (2008)

\bibitem{Bao} Q. Huang, Y. Qiu, W. Bao, J. W. Lynn et al., unpublished,
arXiv:0806. 2776.

\bibitem{Rotter1} M. Rotter, M. Tegel, D. Johrendt, I. Schellenberg, W. Hermes,
and R.Pottgen, Phys. Rev. B \textbf{78}, 020503(R)(2008).

\bibitem{Ni} N. Ni, S. L. Bud'ko, A. Kreyssig, S. Nandi, G. E. Rustan, A. I. Goldman, S. Gupta, J. D. Corbett, A. Kracher, and P. C. Canfield,
Phys. Rev. B \textbf{78}, 014507 (2008).

\bibitem{Ni3} J. Q. Yan, A. Kreyssig, S. Nandi, N. Ni, S. L.
Bud'ko, A. Kracher, R. J. McQueeney, R.W. McCallum, T. A.
Lograsso, A. I. Goldman, and P. C. Canfield, Phys. Rev. B
\textbf{78}, 024516 (2008).

\bibitem{Ni2} N. Ni, S. Nandi, A. Kreyssig, A. I. Goldman, E. D.
Mun, S. L. Bud'ko, and P. C. Canfield, Phys. Rev. B \textbf{78},
014523 (2008).

\bibitem{Tegel} M. Tegel, M. Rotter, V. Weiss, F. M. Schappacher,
R. Poettgen, D. Johrendt, unpublished, arXiv:0806,4782.

\bibitem{Zhao} J. Zhao, W. Ratcliff, J. W. Lynn, G. F. Chen, J. L. Luo, N. L. Wang, J. Hu, and P. Dai, Phys. Rev. B \textbf{78},
140504(R) (2008).

\bibitem{Friedmann} T. A. Friedmann, M. W. Rabin, J. Giapintzakis,
J . P. Rice, and D. M. Ginsberg, Phys. Rev. B \textbf{42}, 6217
(1990).

\bibitem{XHChen3} X. H. Chen, C. H. Wang, G. Y. Wang, X. G. Luo,
J. L. Luo, G. T. Liu, and N. L. Wang, Phys. Rev. B, \textbf{72},
064517 (2005).

\bibitem{Hu} F. Hu, G. T. Liu, J. L. Luo, D. Wu, N. L. Wang, T.
Xiang, Phys. Rev. B \textbf{73}, 212414 (2006).

\bibitem{Kaneko} K. Kaneko, A. Hoser, N. Caroca-Canales, A. Jesche, C. Krellner, O. Stockert, and C. Geibel, unpublished, arXiv: 0807.2608.

\bibitem{Wang} X. F. Wang, T. Wu, G. Wu, H. Chen, Y. L. Xie, J. J. Ying, Y. J. Yan, R. H. Liu, and X. H. Chen, unpublished, arXiv:0806.2452.

\bibitem{Jin} Michael A. McGuire, Andrew D. Christianson, Athena S. Sefat, Brian C. Sales, Mark D. Lumsden, Rongying Jin, E. A. Payzant,
David Mandrus, Yanbing Luan, Veerle Keppens, Vijayalaksmi
Varadarajan, Joseph W. Brill, Raphael P. Hermann, Moulay T.
Sougrati, Fernande Grandjean, and Gary J. Long, Phys. Rev. B
\textbf{78}, 094517 (2008).

\bibitem{Xiang} F. J. Ma, Z. Y. Lu, and T. Xiang, unpublished, arXiv:0806.3526.

\bibitem{WZHu} W. Z. Hu, G. Li, J. Dong, Z. Li, P. Zheng, G. F.
Chen, J. L. Luo, N. L. Wang, unpublished, arXiv:0806.2652.

\bibitem{Chen0}G. F. Chen, Z. Li, G. Li, J. Zhou, D. Wu, J. Dong, W. Z. Hu,
P. Zheng, Z. J. Chen, H. Q. Yuan, J. Singleton, J. L. Luo, and N.
L. Wang, Phys. Rev. Lett. \textbf{101}, 057007 (2008).

\bibitem{WHH} N. R. Werthamer, E. Helfand, and P. C. Hohenberg,
Phys. Rev. \textbf{147}, 295 (1966).

\bibitem{Wen} Y. Jia, P. Cheng, L. Fang, H. Q. Luo, H. Yang, C. Ren, L. Shan, C. Z. Gu, and H. H. Wen, Appl. Phys. Lett. \textbf{93}, 032503 (2008)

\bibitem{Nanda} K. K. Nanda, Physica C \textbf{265}, 26(1996)

\end{thebibliography}

\end{document}